# The Adventure and the Prize

Dennis Sivers
Portland Physics Institute
4730 SW Macadam #101
Portland, OR 97239

Spin Physics Center
Univ. of Michigan 48109
Ann Arbor, MI

ABSTRACT

This article presents a condensed summary of separate, overlapping, presentations to the Workshop on Polarized Drell-Yan Physics (Santa Fe, NM, Oct. 31-Nov. 1, 2010), the GHP 2011 Workshop (Anaheim, CA, Apr. 21-27, 2011) and the Transversity 2011 Workshop (Veli-Lozinj, Croatia, Aug. 29-Sep.2, 2011) during which the author advocated for a potential experimental program based at Fermilab utilizing high intensity polarized proton beams of 120 GeV/c to 150 GeV/c. Some possible experiments in this program are introduced briefly. Interpretations for these potential experiments are presented in terms of a hypothetical solution to the "Quantum Yang-Mills Theory" problem posed by Arthur Jaffe and Edward Witten as one of the seven *Millenium Prize Problems in Mathematics* issued by the Clay Mathematics Institute. These comparisons illustrate the close connection between transverse spin observables and the complex dynamics of confinement and chiral symmetry breaking found in quantum chromodynamics.

## Theoretical Perspective of a Possible Experimental Program at Fermilab With 120-150 GeV/c Polarized Proton Beams

The assortment of theoretical issues surrounding possible experiments at Fermilab utilizing 120-150 GeV/c polarized proton beams of high intensity from the Main Injector can be indicated by briefly considering the official problem "Quantum Yang-Mills Theory" posed by Arthur Jaffe and Edward Witten [1] as one of the seven *Millenium Prize Problems in Mathematics* issued in the year 2000 by the Clay Mathematics Institute. The Jaffe-Witten problem is to demonstrate the existence of a quantum field theory (QFT) based on the non-Abelian gauge symmetry for a compact group, G, in 4-dimensional Minkowski space displaying the properties:

(1) A mass gap, $\Delta$,
(2) Confinement of quarks and gluons,
(3) Chiral symmetry breaking.

A constructive solution to this problem for G=SU(3) would complete the consistent quantum formulation of quantum chromodynamics (QCD) as a theory of the strong interactions. The foundations of QCD include the Nobel-Prize winning work of D.J. Gross, F. Wilczek and H.D. Politzer [2] establishing the property of "asymptotic freedom" that allows a perturbative description of QCD, (PQCD), for dynamical processes involving large momentum transfers, Q. The asymptotic freedom property of QCD is thoroughly established experimentally [3] and the factorization of hard-scattering processes [4] found in PQCD provides for experimental methods to study the short-distance behavior of the standard model and to probe for new dynamical mechanisms beyond the standard model of particle physics [5].

However, PQCD is only a sector of the complete theory. For example, the evolution of $\alpha_S = g_S^2/4\pi$ given by

$$\frac{1}{\alpha_S(Q)} = \frac{1}{\alpha_S(\mu)} + \frac{33-2n_f}{6\pi}\ln\left(\frac{Q}{\mu}\right) \qquad (1.1)$$

when calculated to one-loop accuracy also provides for a long-distance behavior of QCD, when momentum transfers are small compared to the renormalization scale, $\mu$, characterized as "infrared slavery" that is the converse of asymptotic freedom. This infrared behavior points to the importance of an understanding of the strong coupling regime for QCD to provide a QFT description of hadronic physics. To date, no consistent formulation meeting the criteria (1)-(3) above has been found to represent the hadronic sector of the theory. Some reasons why the discovery of such a construction remains an unsolved mathematical challenge are described in the status report on the Jaffe-Witten millenium problem of Michael Douglas [6] while a thorough discussion of empirical "theoretical physics" approaches to understanding the subject can be found in the panel discussion at the Workshop on Quark Confinement and Hadronic Spectrum 9 that is summarized in Ref. [7].

However, physics is an experimental science and nature has already found either a solution to the Jaffe-Witten problem or a way to evade it. Experimental measurements with high-intensity polarized proton beams can provide compelling clues to the significant degrees of freedom for QCD in the strong coupling regime and these clues can guide the construction of an effective field theory. The value of transverse spin observables to the physics issues involved in the Jaffe-Witten problem can be seen clearly if we write the quark terms in the QCD Lagrangian

$$L_q = i\left(\bar{q}_L \gamma^\nu D_\nu q_L + \bar{q}_R \gamma^\nu D_\nu q_R\right) - m_q \left(\bar{q}_L q_R + \bar{q}_R q_L\right) \tag{1.2}$$

in terms of states of definite chirality. For massless quarks the second term on the RHS vanishes and, for $n_f$ massless quark flavors an $SU(n_f)_L \times SU(n_f)_R \times U(1)_L \times U(1)_R$ symmetry is therefore present in the perturbative sector of the theory. The helicity conservation implied by this chiral symmetry is also reflected in the PQCD calculations for transverse spin observables involving light quarks. The result of Kane, Pumplin and Repko [8] published in 1978 (KPR) gives

$$A_N d\sigma(qq\uparrow \Rightarrow qq)/d\sigma(qq \Rightarrow qq) = \frac{\alpha_S(Q)}{Q} m_q f(\theta_{CM}) \tag{1.3}$$

so these observables calculated in PQCD for quark masses [9] such as

$$\begin{aligned} m_u &= 1.9 \pm 0.2 MeV \\ m_d &= 4.6 \pm 0.3 MeV \\ m_s &= 88. \pm 5.0 MeV \end{aligned} \tag{1.4}$$

would be very small. In the mid-80's the KPR result was confronted with the existence of large experimental asymmetries [10,11] in hadronic production processes. The interpretation of the experiments was controversial until it was pointed out [12] that the calculation (1.3) does not, in itself, lead to transverse spin asymmetries for processes involving hadrons containing light quarks. Instead, the KPR result, with $m_q = 0$, implies that PQCD processes can be used to directly probe the soft, dynamical mechanisms that are involved in confinement and chiral symmetry breaking. In the hard-scattering model, these nonpertative mechanisms are factorized, by convention, into hadronic distribution functions or fragmentation functions. When this is done, the description of measurements for transverse single-spin observables pierces straight to the meat of the crucial dynamical issues (1)-(3) that distinguish the hadronic sector of QCD from the short-distance PQCD sector.

To further illustrate the value of transverse spin on this problem, it is helpful to consider some simple quantum symmetries. All single-spin asymmetry measurements of the form,

$$A(\vec{\sigma}) = [M(\vec{\sigma}) - M(-\vec{\sigma})]/[M(\vec{\sigma}) + M(-\vec{\sigma})] \tag{1.5}$$

are odd under an operator, $O$, that acts on a set of 3-vectors, $\vec{k}_i$, and axial 3-vectors, $\vec{\sigma}_j$,

$$O\{\vec{k}_i; \vec{\sigma}_j\}O^{-1} = \{\vec{k}_i; -\vec{\sigma}_j\} \tag{1.6}$$

to serve as a Hodge dual form of the familiar parity operator, P,

$$P\{\vec{k}_i; \vec{\sigma}_j\}P^{-1} = \{-\vec{k}_i; \vec{\sigma}_j\} \tag{1.7}$$

The product of these two operators, $A_\tau = PO$, therefore has the action

$$A_\tau\{\vec{k}_i; \vec{\sigma}_j\} = \{-\vec{k}_i; -\vec{\sigma}_j\}. \tag{1.8}$$

These operators have a group structure defined by $PO = A_\tau, OA_\tau = P, A_\tau P = O$ with $P^2 = O^2 = A_\tau^2 = POA_\tau = 1$ and all can be used to construct idempotent projection operators. This group structure leads to a classification of all single-spin observables into two distinct categories:

1. $P$-odd and $A_\tau$-even
2. $A_\tau$-odd and $P$-even.

In the standard model, $P$-odd, longitudinal single-spin asymmetries are associated with $W_\pm, Z_0$ exchange of the weak interactions while $A_\tau$-odd, transverse single-spin asymmetries involve either quark masses or soft, nonperturbative dynamical interactions in QCD that break the chiral invariance of the quark sector Lagrangian (1.2) [13]. These dynamical mechanisms are precisely the interactions that can be studied at the high-intensity frontier by a fixed-target program with polarized proton beams at the Main Injector.

The result, (1.3) with $m_q = 0$ will be called here KPR factorization. As mentioned above, it implies that the dynamics of transverse single-spin observables can be described in terms of transverse-momentum-dependent distributions and fragmentation functions (TMD's). The application of such functions to hard-scattering processes can be found in the formalism of "TMD factorization" described in the recent book by John Collins [14]. The classification of those $A_\tau$-odd, "leading-twist", TMD distributions and fragmentation functions is based on the work of Mulders and his collaborators [15] while the "Trento Conventions" for relating these quantum structures to the different experimental asymmetries are established in Ref. [16]. These $A_\tau$-odd TMD's consist of two TMD effective distributions (orbital distributions [12], and Boer-Mulders distributions [17]) along with two TMD fragmentation functions (Collins functions [18], and polarizing fragmentation functions). The TMD distribution functions can be called "effective" distributions because the $k_T$-dependence of an orbiting particle in a stable system would vanish by rotational invariance in the absence of initial or final-state interactions to determine that sector of the orbit that is preferentially involved in the scattering process. Such interactions therefore incorporate a "lensing function" [19] into the definition of an $A_\tau$-odd TMD distribution. The requirement for a similar "lensing function" in $A_\tau$-odd fragmentation functions is absent because the orbital angular momentum there occurs explicitly in the final state. The underlying dynamical mechanisms for spin-orbit effects in distributions and fragmentation functions are closely related however, with the real processes involved in jet fragmentation occurring as virtual, $A_\tau$-odd, corrections to the quantum state of a stable hadron. Measurements of TMD's involving transverse spin can thus be an important guide to the strong-coupling limit of QCD.

A field theory description of the hadronic sector of QCD will undoubtedly involve the dynamical degrees of freedom representing **emergent structures** forged from the non-linear dynamics of strongly-coupled quarks and gluons. A partial list of such quantum structures includes:

- constituent quarks
- diquarks
- field-strength densities
- topologically-structured condensates
- virtual mesons and baryons.

With the relative size of the emergent structures being a significant fraction of the size of an individual hadron, a quantitative QFT representation in space-time can be instructive. Sketches of constitutent quarks, scalar and axial vector diquarks pictured in an Abelian coordinate gauge are shown in Fig. 1. These complicated structures surround the quark partons of PQCD and clues to the soft dynamics can be found in hard-scattering processes.

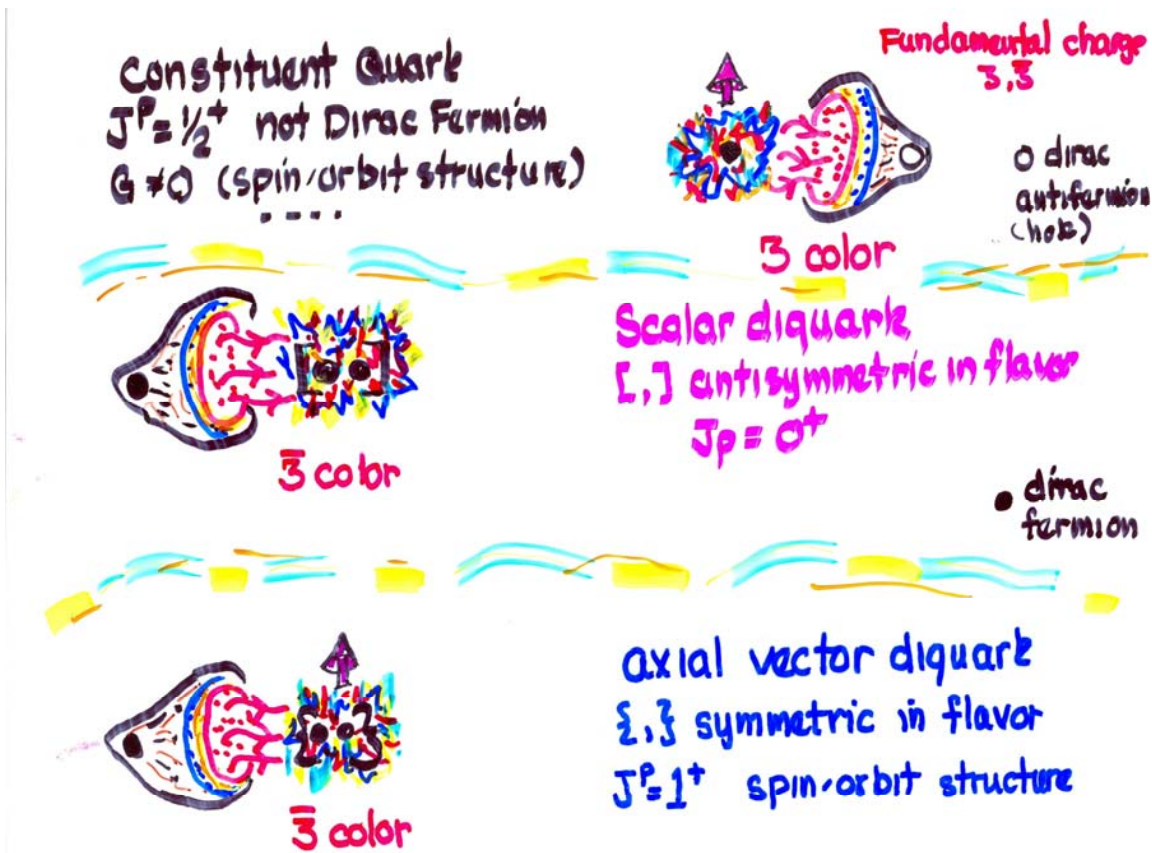

Fig. 1. Sketches indicating the structure of constituent quarks and of scalar and axial vector diquarks in an Abelian coordinate gauge.

The spatial representation of adjoint charges is also important in understanding the confinement of color. Field strength densities with scalar and pseudoscalar quantum numbers are indicated in Fig. 2. Lattice-gauge theory calculations [20] provide tools to study these degrees of freedom in stable hadrons and lattice techniques certainly provide the most important approach to the study of nonperturbative dynamics in QCD. However, lattice regularization methods in four-dimensional Euclidean space are not suited to study the dynamics of highly virtual systems that appear in jet fragmentation. This is one area where the dynamical information from transverse single spin asymmetries can prove particularly useful in formulating an effective field theory approach to hadronic structure. The quantitative comparison of asymmetries involving hadronic distributions to those formed from the fragmentation process offers great opportunities to study crucial facets on nonperturbative QCD in new ways.

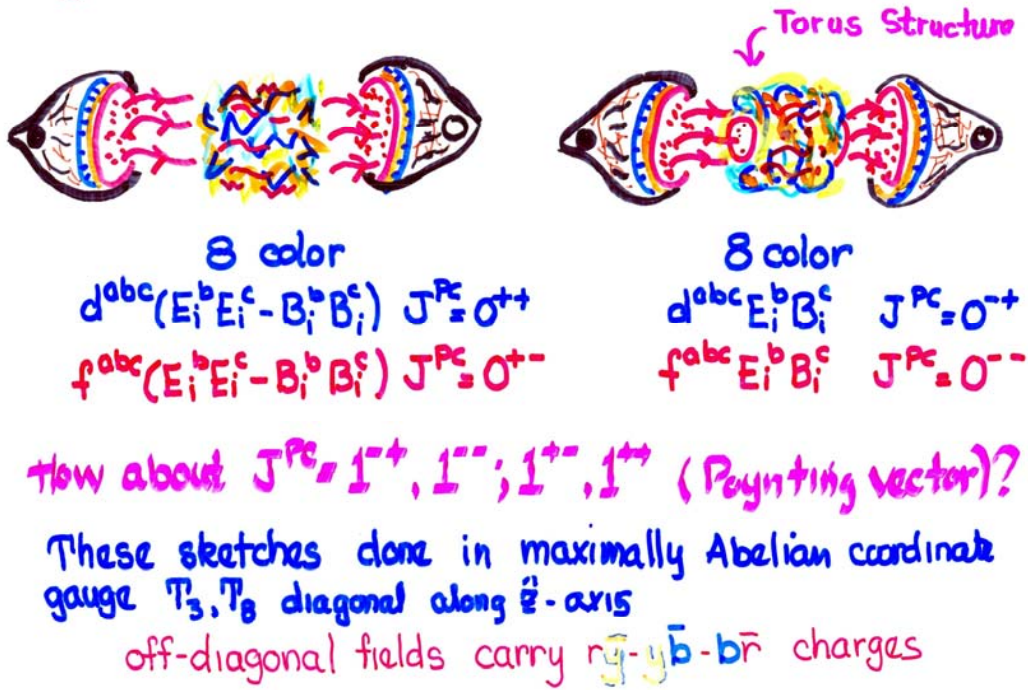

Fig. 2 Spatial structure of non-Abelian field strength densities with adjoint color charge are shown in an Abelian coordinate gauge.

Specific experimental measurements with polarized proton beams can provide different types of information. We therefore illustrate the theoretical issues relating transverse spin observables to confinement and chiral symmetry breaking in QCD with three sample experiment types in which a fixed-target program involving polarized proton beams can have a significant impact. The first involves the Drell-Yan process.[21] The two-spin asymmetry in the Drell-Yan process, $A^{NN} d\sigma[p\uparrow p\uparrow \Rightarrow (l^+l^-)X]$, can provide direct information on the transversity distributions, $\delta_N q^i(x,\mu^2)$, for quarks and, $\delta_N \bar{q}^i(x,\mu^2)$, for antiquarks in a transversely polarized proton. One way to appreciate this information is to consider the comparison of the transversity distributions to the helicity distributions, $\Delta_L q^i(x,\mu^2)$ and $\Delta_L \bar{q}^i(x,\mu^2)$, indicated schematically in Fig. 3. Since the $L_z$ quantum number is preserved under boosts along the z-direction and only the s-wave component of the spin-1/2 partons in a $J=1/2$ system can contribute to the $k_T$-integrated probability distributions of the rotated system, we can write the distributions

$$\Delta_L q^i(x,\mu^2) = \sum_{L=-\infty}^{L=+\infty}(|a_L^i|^2 - |b_L^i|^2)$$
$$\delta_N q^i(x,\mu^2) = (|a_0^i|^2 - |b_0^i|^2). \qquad (1.9)$$

This means that the difference between the helicity distributions and the transversity distributions for quarks and antiquarks provides a direct, x-dependent, measure of nonzero orbital angular momentum components in the wave function of the proton. Extraction of the transversity distributions for u,d quarks from global fits [22] to data on the Collins-Heppelman asymmetry [23] in SIDIS combined with fits to the Collins functions extracted from $e^+e^-$ asymmetries [24], has already provided considerable information about proton orbital structure.

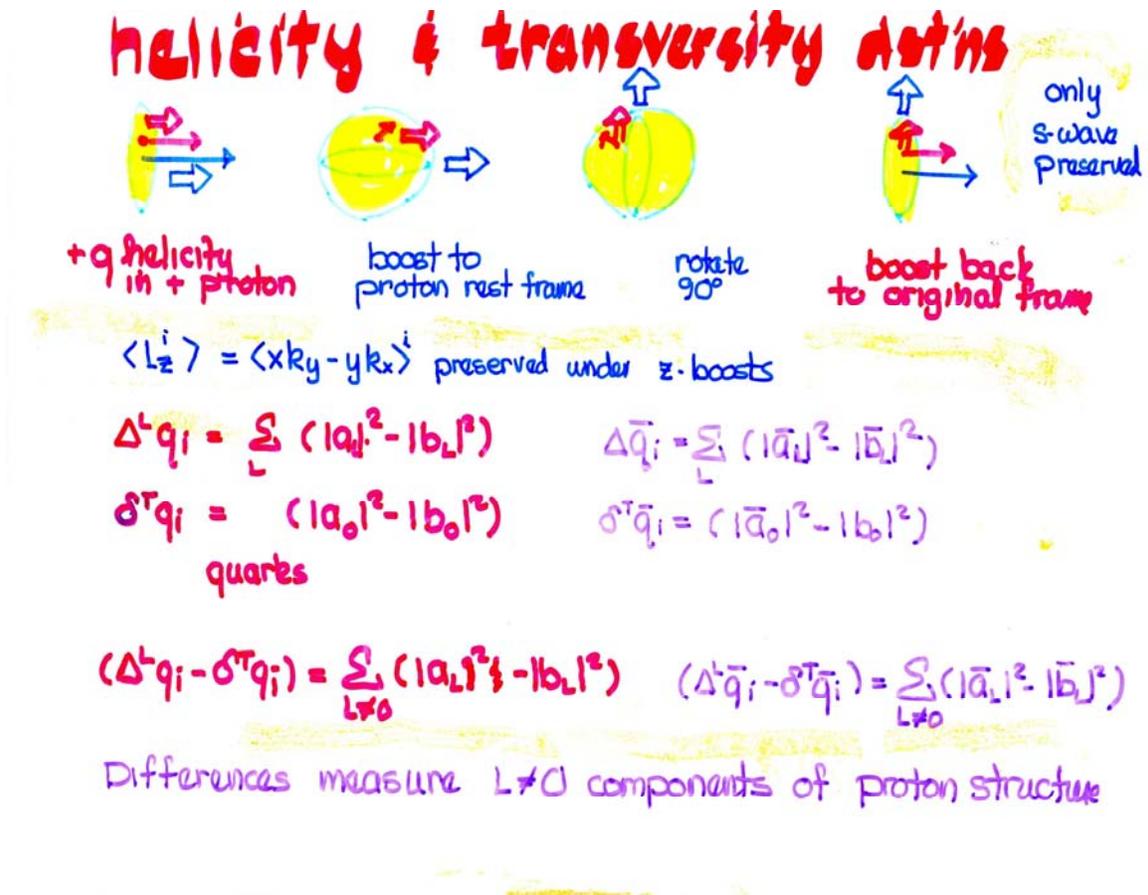

Fig. 3. These Sketches indicate the relationship between transversity and helicity distributions.

A nonzero value of the Drell-Yan two-spin asymmetry in proton proton collisions can therefore be used to measure the transversity distributions for antiquarks, and

$$\delta_N \bar{q}^i(x,\mu^2) = \Delta_L \bar{q}^i(x,\mu^2) - \sum_{L\neq 0}(|\bar{a}_L^i|^2 - |\bar{b}_L^i|^2) \qquad (1.10)$$

can thus provide important information about the S-wave component of the spin-polarized sea.

The process dependence of $A_\tau$-odd effective distributions provides a very unique and non-intuitive test of the connection of emergent hadronic structures to the underlying gauge theory. Based on the gauge-link formalism for nonlocal quark correlators, John Collins [25] found the relationship

$$\Delta^N f_{q/p\uparrow}^{DY}(x,k_T;\mu^2) = -\Delta^N f_{q/p\uparrow}^{SIDIS}(x,k_T;\mu^2). \tag{1.11}$$

This gives the connection between the orbital distribution measured in the Drell-Yan process (DY) and the orbital distribution measured in semi-inclusive deep inelastic scattering. F. Pijlman [26] has noted that the path integral approach to the respective nonlocal correlators places the result (1.11) in a formal analogy to the calculation of the Aharonov Bohm [27] asymmetry in QED.

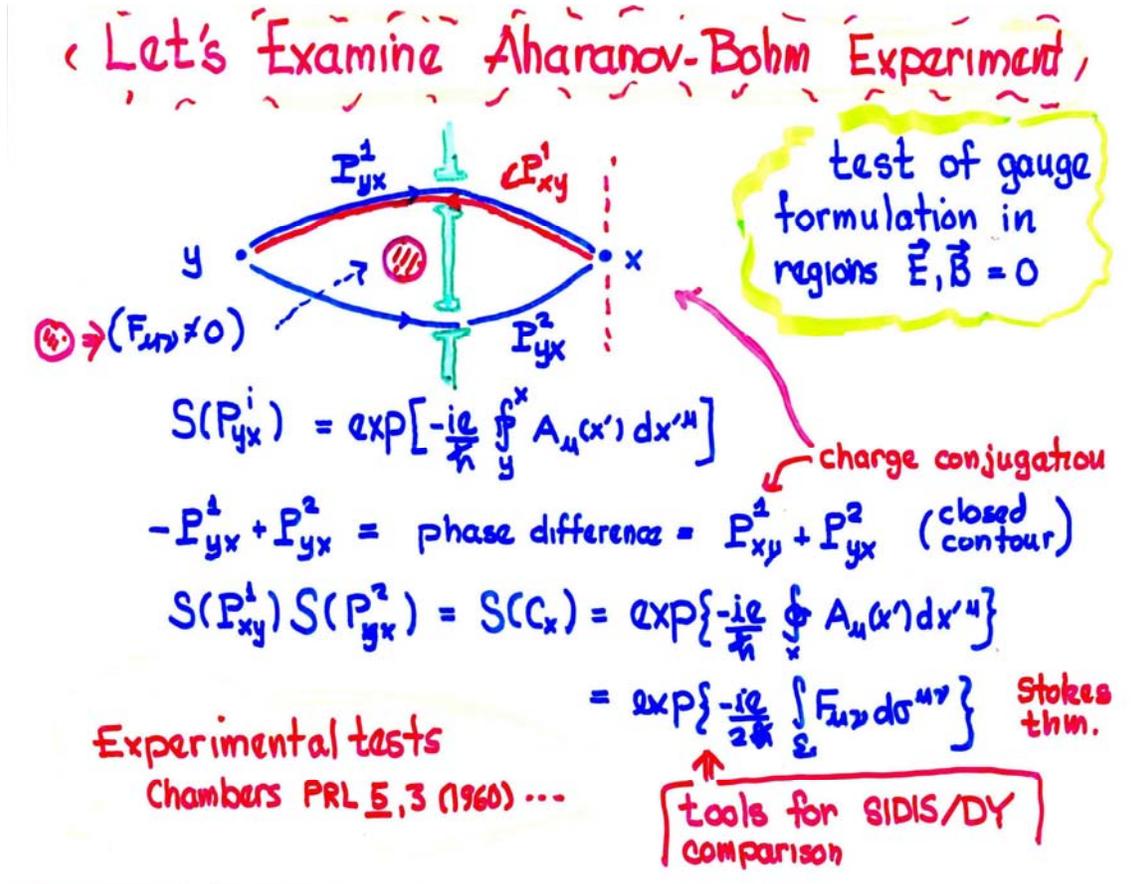

Fig. 4 The Aharanov Bohm asymmetry for charged beams.

Whereas the Aharanov Bohm results sketched in Fig. 4 test QED gauge invariance in spatial regions where the electromagnetic field-strength tensor vanishes the Collins prediction, (1.11), indicated in Fig. 5 tests gauge invariance for QCD in a regime where complicated nonlinear dynamics create quantum structures that are quite different from the partons of PQCD. Since the goal is, not only to observe a sign change, but to measure the equality of the magnitude of the asymmetries, a quantitative experimental study of (1.11) can be greatly aided by measurements of the DY process at the **high-intensity frontier** of polarized proton beams.

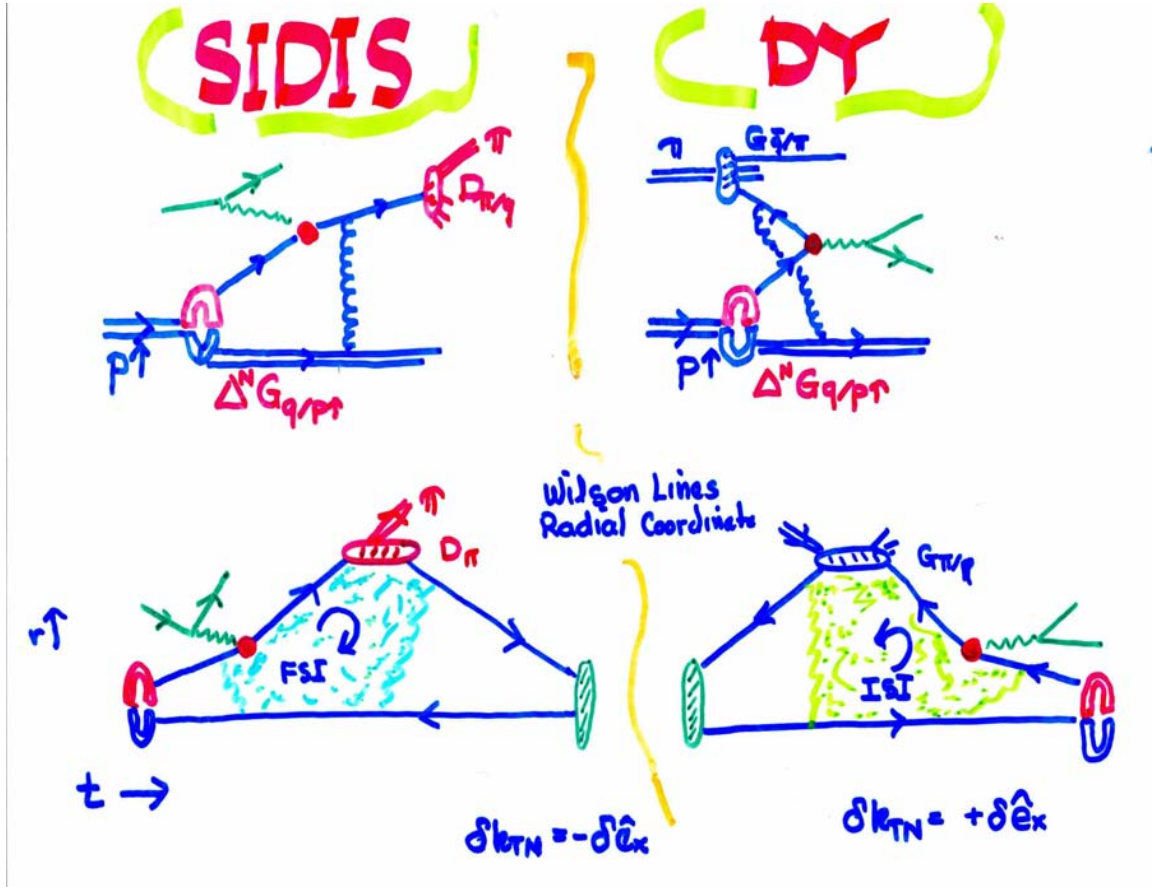

Fig. 5 Wilson lines for Collins conjugation involving arbitrary soft interactions are compared to simple one-gluon spectator models for single-spin asymmetries in SIDIS and DY.

Experimental verification of the Collins conjugation result would certainly provide substantial positive support for the conjectured existence of a QFT formulation of hadronic physics based on an underlying SU(3) gauge symmetry. In contrast, an experimental contradiction with (1.11) could indicate the need for separate dynamic degrees of freedom—not related to the gauge symmetries of PQCD—that are required to understand the full range of hadronic interactions. That would then suggest that the Jaffe-Witten millenium mathematics challenge remains, in some sense, incompletely formulated so that no existence proof is possible without additional assumptions.

A different kind of information that emphasizes the importance of observing chiral symmetry breaking both in hadronic distributions and in the fragmentation process can be found from measurement of inclusive asymmetries for the production of one or more mesons. In the semi-inclusive process

$$A^N d\sigma(ep\uparrow \Rightarrow e\pi^+ X) = \frac{1}{2}[d\sigma(ep\uparrow \Rightarrow e\pi^+ X) - d\sigma(ep\downarrow \Rightarrow e\pi^+ X)],$$ it is possible to

distinguish kinematically the $A_\tau$-odd asymmetries generated by spin-orbit dynamics in the proton from those asymmetries produced in the Heppelman-Collins-Ladinski (HCL) mechanism where the $A_\tau$-odd dynamics appears in the Collins fragmentation functions. For inclusive single spin asymmetries involving meson production for polarized proton beams such as $A^N d\sigma(pp\uparrow \Rightarrow \pi^+ X)$ and $A^N d\sigma(pp\uparrow \Rightarrow \pi^- X)$, it is not possible to make such a separation. Phenomenological studies [28] of the large asymmetries for such processes found at Fermilab [29] and RHIC [30] have to include contributions from both types of mechanism

$$A^N d\sigma(pp\uparrow \Rightarrow \pi^+ X) \cong A^N_{HCL} d\sigma(pp\uparrow \Rightarrow \pi^+ X) + A^N_{orbit} d\sigma(pp\uparrow \Rightarrow \pi^+ X) \quad (1.12)$$

to describe the data. The first term in (1.12), $A^N_{HCL}$, can be calculated from the quark transversity distributions, the Collins functions and the perturbatively-calculable quark spin transmission parameters, $C_{N0;N0}$, for quark-quark and quark-gluon scattering. The second term, $A^N_{orbit}$, however, involves unknown lensing functions for the "effective" orbital distributions functions [31]. Since the connection between jet structure and hadronic structure can provide important information about the underlying QFT, it is interesting to connect the measurements (1.12) with asymmetries in other processes in which the two mechanisms can be clearly separated.

An example of a process that is uniquely appropriate for a fixed-target program with high-intensity polarized proton beams in which distribution effects can be experimentally separated from fragmentation effects can be seen in Fig. 6. These sketches indicate a measurement of

$$A^N d\sigma(pp\uparrow \Rightarrow \pi^+\pi^- X) \cong A^N_{HCL} d\sigma(pp\uparrow \Rightarrow \pi^+\pi^- X) + A^N_{orbit} d\sigma(pp\uparrow \Rightarrow \pi^+\pi^- X) \quad (1.13)$$

for nonresonant pion pairs in the same jet presents a situation in which spin-orbit dynamics in the distribution function and spin-orbit dynamics in fragmentation can be distinguished experimentally. This is an example of the application of the property of KPR factorization [32] to the isolation of a spin-asymmetry defined in terms of a spin-directed momentum transfer of the form $\delta k_{TN}(h) = \delta[\vec{k}_h \cdot (\hat{s} \times \hat{P}_{jet})]$.

$$A^N_{HCL} : \langle \delta k_{TN}(\pi^+) + \delta k_{TN}(\pi^-) \rangle \to 0; \langle \delta k_{TN}(\pi^+) - \delta k_{TN}(\pi^-) \rangle \neq 0 \quad (1.14)$$

$$A^N_{orbit} : \langle \delta k_{TN}(\pi^+) + \delta k_{TN}(\pi^-) \rangle \neq 0; \langle \delta k_{TN}(\pi^+) - \delta k_{TN}(\pi^-) \rangle \to 0 \quad (1.15)$$

The point here is to distinguish $A_\tau$-odd effects from $A_\tau$-even dynamics in the fragmentation process by looking at two particle distributions.

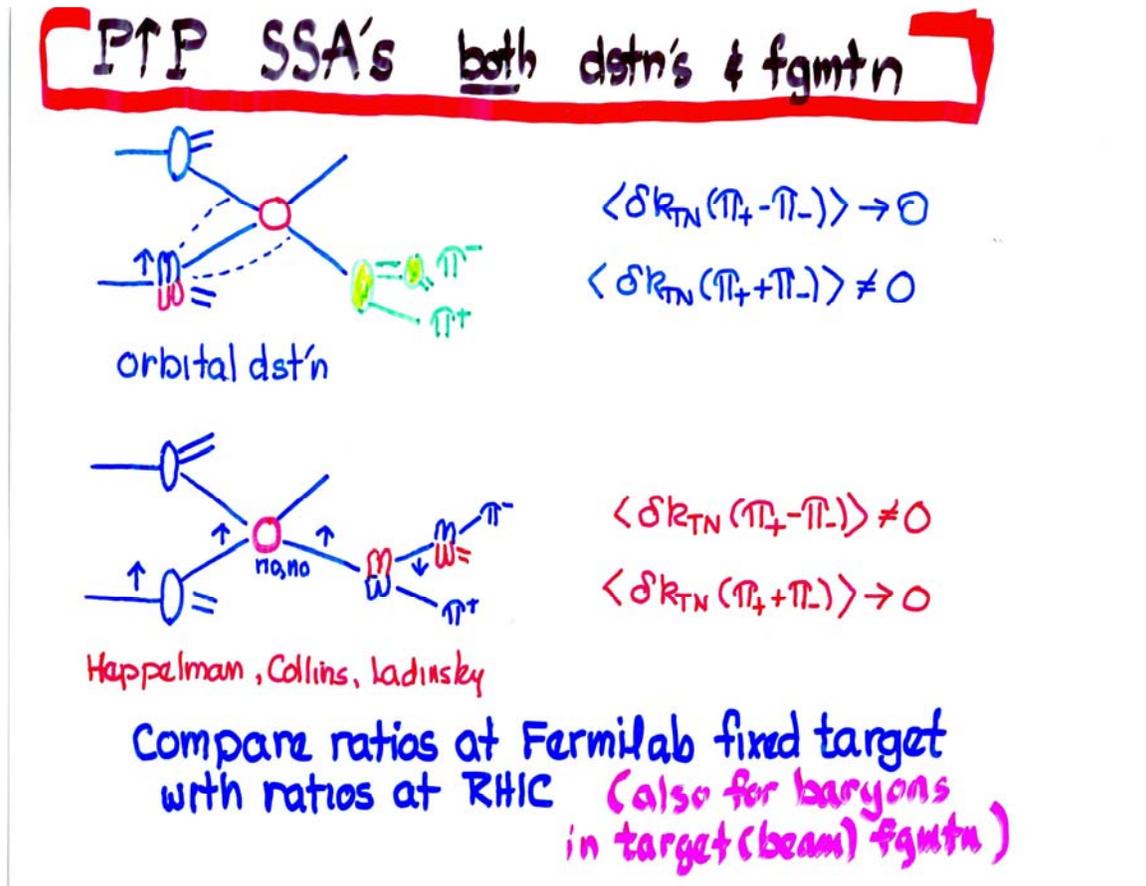

Fig. 6 The application of the concept of spin-directed momentum can isolate $A_\tau$-odd dynamics into either hadron distribution functions or into the fragmentation process.

Another way to see how the concept of using spin-directed momentum to isolate $A_\tau$-odd dynamics can lead to identifying completely new types of spin asymmetries can be found in the correlation of particle production in jets with the orientation of hyperon spins as discussed in Ref. [33].

The final type of experiment we intend to discuss as a guide to confinement and chiral symmetry breaking in hadronic physics involves one-spin and two-spin asymmetries in elastic proton proton scattering. Such experiments [34] have presented some of the earliest indications for the importance of transverse spin observables. To describe the connection to a QFT for hadronic physics, we will review briefly the basis for an effective field theory for exclusive scattering processes at fixed scattering angle. Based on the work by Matveev et al. [35] and by Brodsky and Farrar [36] it is possible to argue that all fixed-angle exclusive hadronic scattering processes should have the asymptotic form

$$\lim_{s\to\infty}\frac{d\sigma_{ab\Rightarrow cd}(s,\theta_{CM})}{dt}=\left(\frac{\langle m^2\rangle}{s}\right)^{n_a+n_b+n_c+n_d-2}f_{ab\Rightarrow cd}(\theta_{CM},\alpha_S) \quad (1.16)$$

at fixed CM angle. In this expression $n_a$ is the minimal number of partons in hadron a, and so on. The use of the term "parton" instead of "constitutent" in labeling the appropriate degrees of freedom is justified by the sophisticated geometric arguments leading to the counting rule result. The component of each hadronic wave function leading to the asymptotic behavior (1.16) is one in which the valence "constituents" of the hadrons overlap within a radius $R_C\approx\frac{c_{\text{int}}}{\sqrt{s}}$ and the quantum fluctuations displayed in Figs. 1-2 are consequently suppressed. The overlap of all valence partons from incoming and outgoing hadrons creates a small, highly virtual, system in which PQCD can be applied. Consequently, Efremov and Ryadushin [37] and Brodsky and Lepage [38] (BLER) separately showed that, in this limit, hadronic scattering amplitudes can be expressed in the factorized form

$$\lim_{s\to\infty}\langle\lambda_c\lambda_d|M|\lambda_a\lambda_b\rangle=\sum_{\lambda_i}\int\prod[dx_i]\phi^*_{c\lambda_c}(x_{ci},\lambda_i)\phi^*_{d\lambda_d}(x_{di},\lambda_i)T_H(x_i,\lambda_i;s,t,\alpha_S)\phi_{a\lambda_a}(x_{ai},\lambda_i)\phi_{b\lambda_b}(x_{bi},\lambda_i)$$
(1.17)

in terms of hadronic wave functions, $\phi_{b\lambda_b}(x_{ai},\lambda_i)$, and a multi-parton hard-scattering amplitude, $T_H(x_i,\lambda_i;s,t,\alpha_S)$, that is calculable in PQCD. In Eq. (1.17), the hadron helicities $\lambda_a-\lambda_d$ are explicitly displayed along with the parton helicities $\lambda_i$. We are applying the formula for light quarks, $m_q\to 0$, where quark helicities are preserved and there are no transverse spin observables generated in the PQCD hard scattering amplitude. The direct application of this formula to meson-baryon and baryon-baryon scattering amplitudes involves several challenges. The cogent issues are summarized in the work of Isgur and Llewelyn Smith [39]. In spite of the fact that there are approximately 300,000 distinct tree graphs in a Feynman diagram approach to $6q\Rightarrow 6q$, it is still possible to make progress [40] in understanding $NN\Rightarrow NN$ elastic scattering amplitudes in the Jacob-Wick [41] helicity formalism. The amplitudes are conventionally defined as

$$\begin{aligned}\Phi_1(s,\theta)&=\langle++|M|++\rangle\\ \Phi_2(s,\theta)&=\langle++|M|--\rangle\\ \Phi_3(s,\theta)&=\langle+-|M|+-\rangle\\ \Phi_4(s,\theta)&=\langle+-|M|-+\rangle\\ \Phi_5(s,\theta)&=\langle++|M|+-\rangle.\end{aligned} \quad (1.18)$$

The connection of the BLER formalism for hadron-hadron exclusive processes to transverse spin observables and a QFT description of hadronic physics requires a discussion of the Landshoff process [42] and a generaliztion of the original collinear formulation of (1.17) to introduce transverse structure (either partonic impact parameters or transverse momenta) into the distribution amplitudes, $\phi_{a\lambda a}$. The geometrical analysis that provides the basis to compare the Landshoff mechanism with the counting-rule diagrams shown in Fig. 7 can be found in the excellent review article on fixed-angle exclusive hadron scattering processes by Sterman [43].

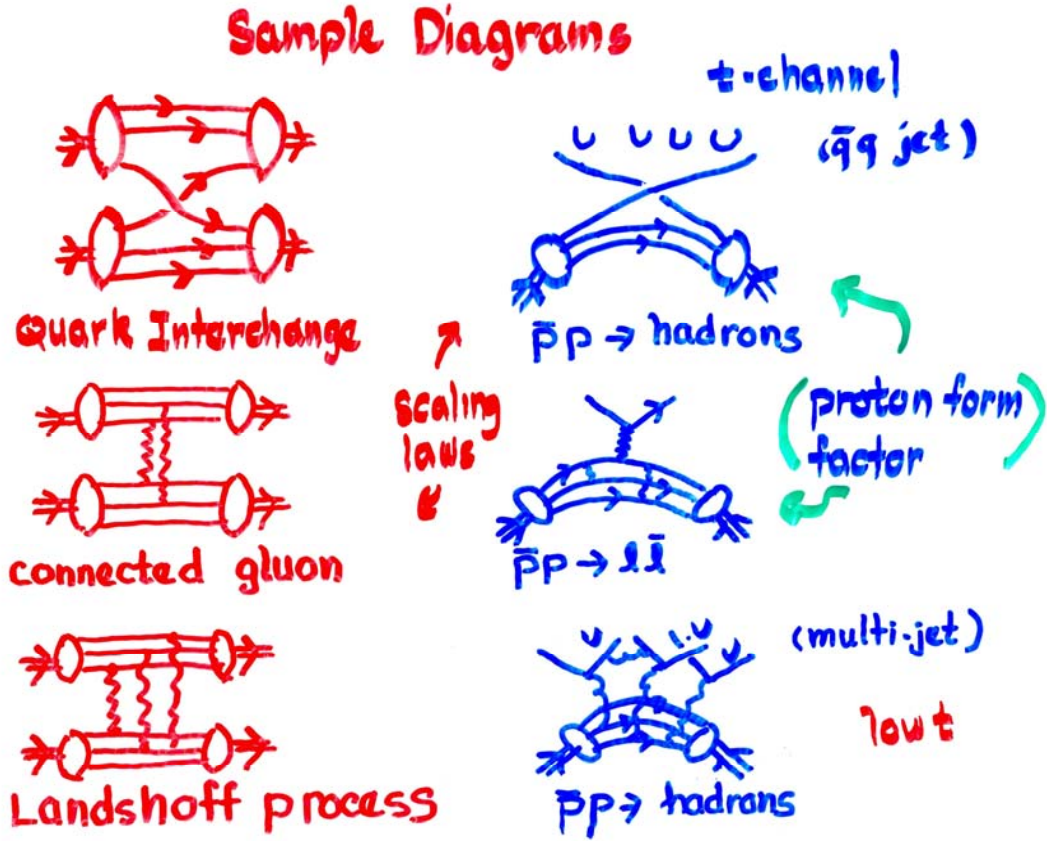

Fig. 7 Sample diagrams showing the classes of diagrams leading to fixed-angle proton-proton elastic scattering.

The basic Landshoff diagram for proton-proton elastic scattering shown here involves 3 independent qq scattering processes that overlap to lead to an elastic cross section

$$\lim_{s\to\infty}\frac{d\sigma_{pp\Rightarrow pp}(s,t)}{dt} = \left(\frac{\langle m^2\rangle}{t}\right)^8 F_L(s,\alpha_S) \qquad (1.19)$$

Unlike the "connected" diagrams in Fig. 7 where radiative corrections are small because of the small size of color-singlet clusters, the disconnected Landshoff diagrams are found in PQCD calculations [44] to be suppressed by the resummation of perturbative diagrams that lead to Sudhakov [45] factors,

$$S_{qq}(t,t_0;\alpha_S) \cong \exp[-\frac{\alpha_S}{3\pi}\log^2(t/t_0)] \qquad (1.20)$$

The comparison of Landshoff diagrams to connected diagrams introduces the concept of transverse structure of the proton into the calculation of the amplitudes. It is interesting to consider the ratio of the Landshoff contribution to one of the helicity-conserving amplitudes in Eq. (1.18) to the contribution from the connected diagrams. Based on the calculations of Mueller [46] and of Botts and Sterman [47], the ratio can be approximated

$$\frac{\Phi_1^L(s,\theta)}{\Phi_1^Q(s,\theta)} \approx \frac{c_1^L}{c_1^Q}(s/s_0)^{\gamma(\theta,\alpha_S)} \tag{1.21}$$

This calculation involves a competition between a geometric enhancement and a radiative suppression for the Landshoff process compared to the counting-rule amplitudes. At fixed t ($\theta = 0$) we have $\gamma(0,\alpha_S) \cong 1$ and the Landshoff mechanism dominates. For $\theta \neq 0$, $\gamma(\theta,\alpha_S)$ decreases significantly but the ratio (1.21) remains uncertain at large angles. A significant portion of this uncertainty is directly related to the mass-gap, $\Delta$, in the Jaffe-Witten problem. Consider a color-singlet cluster of 3 quarks and $N_G$ gluons with an invariant mass squared, $M_{cl}^2$, that is isolated from other particles. Because of the absence of a mass gap in PQCD calculations, the question of whether this cluster describes a virtual proton contributing to the elastic cross section or whether it represents an inelastic event depends on an arbitrary cutoff. In the full quantum theory of QCD, assuming the existence of a mass gap, $\Delta = m_\pi$, an inelastic event involves the creation of, at least, one additional particle with a threshold given by

$$M_{cl}^2 \geq M_{th}^2 = m_p^2 + 2m_p m_\pi + m_\pi^2 \tag{1.22}$$

The matching of calculations in PQCD with hadronic cross sections in inclusive processes involves the formulation of IR-safe observables formed by averaging over ensembles involving many hadrons. This approach is not available in the calculation of exclusive amplitudes and the complicated, nonperturbative effects giving the cut-offs must be hidden in the hadronic wave functions, $\phi_{a\lambda}(x_{ai},\lambda_i)$, of Eq. (1.17).

The helicity non-conserving amplitudes, $\Phi_5(s,\theta)$ and $\Phi_2(s,\theta)$, given in (1.18), also provide an important measure of transverse proton structure. Using the KPR limit of quark helicity conservation combined with parity and time-reflection invariance, these amplitudes can be written in terms of the helicity conserving amplitudes. [48]

$$\Phi_5(s,\theta) \cong [\varepsilon(t) - \varepsilon(u)]\Phi_1(s,\theta) + [\varepsilon(t) + \varepsilon(u)][\Phi_3(s,\theta) + \Phi_4(s,\theta)]$$
$$\Phi_2(s,\theta) \cong \frac{1}{2}[\varepsilon(t) - \varepsilon(u)]\Phi_5(s,\theta) \tag{1.23}$$

In this expression $\varepsilon(t)$ is a spin-flip factor,

$$\varepsilon(t) \cong \frac{\varepsilon_0 (-t)^{\frac{1}{2}}}{m_p^2 - t}, \tag{1.24}$$

that represents the overlap of an amplitude where the proton helicity in either the initial or final states does not match the sum of the quark helicities. The constant, $\varepsilon_0$, is a small complex number. Taking $|\varepsilon_0|$ as an expansion parameter and systematically discarding terms of $O(|\varepsilon_0|^2)$ gives the full set of elastic observables in terms of the 3 helicity conserving amplitudes and the constant, $\varepsilon_0$. Let $\Sigma = \frac{1}{\pi}(s-4m_p^2)d\sigma/dt$, we have

$$\begin{aligned}
\Sigma &= \frac{1}{2}(|\Phi_1|^2 + |\Phi_3|^2 + |\Phi_4|^2) \\
P\Sigma &= \text{Im}[\Phi_5^*(\Phi_1 + \Phi_3 - \Phi_4)] \\
A_{SL}\Sigma &= \text{Re}[\Phi_5^*(\Phi_1 - \Phi_3 + \Phi_4)] \\
A_{NN}\Sigma &= -\text{Re}(\Phi_3\Phi_4^*) \\
A_{SS}\Sigma &= \text{Re}(\Phi_3\Phi_4^*) \\
A_{LL}\Sigma &= \frac{1}{2}(-|\Phi_1|^2 + |\Phi_3|^2 + |\Phi_4|^2).
\end{aligned} \quad (1.25)$$

A nontrivial test of these approximations is found in the inequalities,

$$|A_{NN} + A_{SS}| \le |P| : |A_{NN} + A_{SS}| \le |A_{SL}| \quad (1.26)$$

A phenomenological fit to the helicity conserving amplitudes, $\Phi_1, \Phi_3, \Phi_4$, in terms of Regge components, Landshoff, Connected Gluon and Quark Interchange diagrams combined with the spin-flip factors (1.23) and (1.24) provides a very economical way to describe the data. The underlying angular dependence and the flavor dependence for all the amplitudes can be found by matching the classes of Feynman diagrams, in either the u-channel or the t-channel, to similar diagrams found in the calculation of the proton form factor. The remaining uncertainty involves the relative normalization of the Landshoff contribution with respect to the counting rule diagrams as indicated above in Eq. (1.21).

The location of the existing data [34] on $A_{NN}$ and $P$ for elastic $pp \Rightarrow pp$ is shown on the Mandelstam plane in Fig. 8. Also indicated in this drawing are rough estimates for the region of validity of the approximations described above.

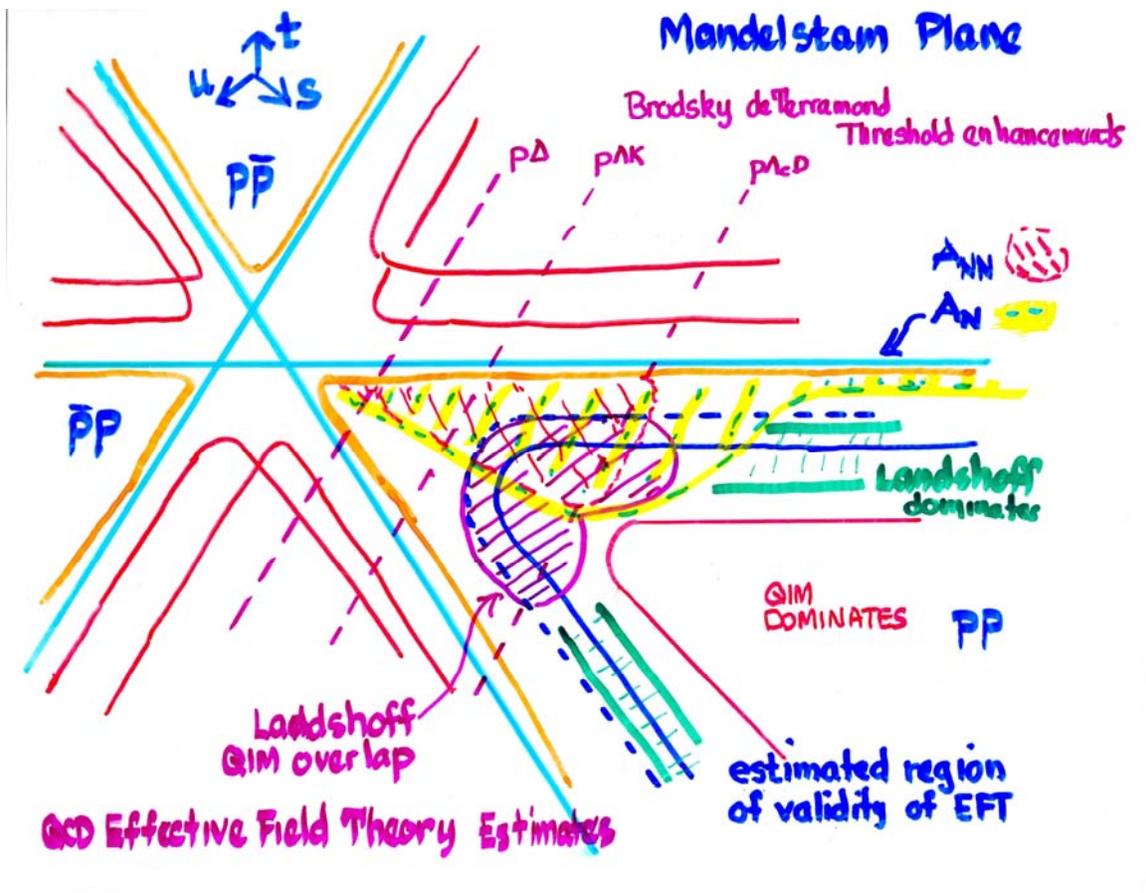

Fig. 8 Existing data on spin observables for elastic pp scattering and the estimated region of validity for fits.

Fits to the existing data on $A_{NN}$ have been performed by Pire and Ralston [49] and by Botts [50] based on the Landshoff and quark interchange mechanisms. Ramsey and Sivers [48,51] have also done such fits but have included connected gluon amplitudes to fit the ratio of the $np \Rightarrow np$ cross section to the $pp \Rightarrow pp$ cross section and have also fit the polarization data based on Eq. (1.23). While differing in details, all these approaches find evidence for significant interference effects between the Landshoff and quark interchange mechanisms implied by the structure in $A_{NN}$ observed experimentally. In contrast, Brodsky and de Teramond [52] have argued for a much stronger suppression of the Landshoff mechanism than that found by Mueller or Botts and Sterman. Instead they advocate attributing the structure in $A_{NN}$ found in Refs. [48-51] to the interference between quark interchange amplitudes and a $J = 1$ dibaryon resonance at $\sqrt{s} = 5.08$ GeV representing a threshold enhancement, $pp \Rightarrow p\Lambda_c D$, associated with charm production. Partial support for this innovative idea can be found in the connection between another enhancement in the data and the $pp \Rightarrow p\Lambda K$ threshold shown in Fig. 8.

Whatever conclusions that can be drawn by testing these ideas against existing data, the discovery potential for new information on chiral dynamics and hadronic structure from additional experiments on proton-proton spin observables at large transverse momentum is rated to be very high. Obviously, high-intensity polarized proton beams are a key requirement for this type of measurement.

We have now used three different types of experiment to illustrate some common underlying features of nonperturbative dynamics in QCD. The discussion has been primarily designed to remind those physicists interested in the topic of transverse spin physics, either as active participants or informed observers, of the close connection have with the ultimate goal of understanding color confinement and chiral dynamics in QCD within the context of a consistent quantum field theory. The references to the "Quantum Yang-Mills" problem posed by Arthur Jaffe and Edward Witten made in the discussion of the experiments were done with the understanding that the Millenium Prize problem remains, solely, a problem in mathematics. If, or when, the problem is solved it will be solved within the magnificent framework of rigid logic demanded my mathematical proof. Whether or not such a proof would immediately provide tools for the calculation of hadronic cross sections cannot yet be known. The achievement would certainly provide inspiration to theoretical physicists. Meanwhile, the empirical study of hadronic structure, whether viewed through transverse-momentum dependent distribution functions and fragmentation functions, through generalized parton distributions, light cone wave functions or Wigner distributions, currently represents a constructive discipline of using what is known about PQCD to study the remaining mysteries of QCD itself. While mathematicians and theoretical physicists share similar goals, they have significantly different approaches to the subject. Experimental measurements, however, will ultimately determine the scope of our fundamental understanding of hadronic physics. The specific goal of creating a fixed-target experimental program at Fermilab that can be part of this adventure is more fully described in the Updated Report: Acceleration of Polarized Protons to 120-150 GeV/c at Fermilab. [53]